# Static analysis-guided agentic AI translation enables Rust as a full stack bioinformatics language

Johan Henriksson[1,2,3,4]

[1] Department of Molecular Biology, Umeå University, Umeå, Sweden
[2] Umeå Centre for Microbial Research (UCMR), Umeå University, Umeå, Sweden
[3] Integrated Science Lab (IceLab), Umeå University, Umeå, Sweden
[4] Science for Life Laboratory (SciLifeLab), Umeå University, Umeå, Sweden

**The field of bioinformatics struggles with legacy code - old code that is commonly used but may no longer have a maintainer, or may be written in an now-unfamiliar language (e.g. Perl, Fortran). This incurs maintenance cost (technical debt), but dynamically typed languages also negatively impacts the environment and fail to make use of modern hardware. Legacy code may also have security or safety problems that make it unsuited for use in clinical settings. Here we show that agentic AI, combined with static analysis, can be used to translate legacy code to the modern language Rust. We provide prompts and supporting software to aid systematic translation, and evaluate it on common software for NGS and imaging. We showcase the result on our software Bascet: Size was reduced by ~80x, build time decreased by ~10x, and performance of key steps improved >3x. Unix dependencies were also removed, making Bascet the only single-cell pipeline able to run on native Windows, without a container. Large-scale refactoring of bioinformatics software is thus now possible at a limited budget, enabling more complex tools to be developed.**

# Introduction

Bioinformatics software underpins all major biological studies today, yet funding for software is scarce, the need for software maintenance is not given enough attention[1], and there is a lack of workforce[2]. While major version releases can be published anew, bug fixes typically cannot, and there is commonly little incentive for a non-author to contribute to someone else's software. This has resulted in a fragmented bioinformatics ecosystem, where it can be hard to combine software (R vs Python[3]), be hard to run on modern computers (requiring outdated libraries), not use modern hardware efficiently, or it may no longer scale for newer large-scale datasets. When software is poorly designed (commonly due to budget or time constraints, PhDs or postdocs leaving the lab, lack of training, or change of design), issues arise that need to be circumvented when the software is used later. This is referred to as *technical debt* - money saved upon software conception has to be paid for later maintenance when the software is actually used (e.g., by setting up dedicated software environments or fixing the bugs). Research software has particularly high technical debt[4], with issues taking a long time to resolve[5]. New challenges are now also on the horizon: (1) R, Python and Matlab consume >50x more energy than compiler or semi-compiled languages for certain workloads[6], making them bad for the environment. (2) Hardware is becoming scarce due to ramped up demand, requiring code to be more efficient in the coming years. (3) Computer hardware is evolving away from the use of CPUs to more efficient architectures, requiring extensive software rewriting[7]. All of these major problems call for a rethinking of how bioinformatics research is being conducted.

Strong initiatives for improving bioinformatics code do exist; e.g., Bioconductor is an effort to increase the interoperability and quality of R-packages[8]. While the same software may produce different results when run on different operating systems, Nextflow has narrowed this gap[9]. However, despite the best efforts, major problems persist: It can be difficult to mix R and Python code, and the choice of language is largely based on package availability - i.e., RNA-seq analysis is almost synonymous with R[3] while machine learning is synonymous with Python. Similarly, the library Bioformats is a cornerstone for microscopy[10], supporting 165 file formats (v8.5), but being written in Java, requires use of the Java virtual machine, limiting interoperability. Bioinformaticians are thus expected to learn at least two programming languages to cover most use cases. However, the increased demand for web applications also adds the need for Javascript, and high performance computing further requires knowledge of any of C, C++, Rust, Golang, Zig, or CUDA/OpenCL in the extreme case. The need to learn deeply about such a large range of topics is incompatible with the pressure to publish[11], and progress quickly in academia[11–13].

In this study we show how agentic AI, combined with classic static analysis, has the potential to rewrite all bioinformatics software in a single modern language. To be able to handle all use cases, this eliminates R and Python as the one language (not suited for implementing, e.g., aligners). Also, C/C++ are generally too difficult to use for most bioinformaticians due to poor tooling (e.g., no standardized build system nor docucumentation system), legacy code dependencies and unsafe language features (pointers). However, newer languages now exist. Rust was designed by Mozilla foundation to handle the requirement for performance and security in their Firefox webbrowser (https://rust-lang.org/). It has since been accepted in the Linux kernel, and since C was made popular due to Unix, Rust has by analogy the potential of becoming the next dominant language. Major companies such as Dropbox[14], Discord[15], Vercel[16], Cloudflare[17], Meta/Whatsapp[18] have already migrated significant parts of their code to Rust, both to improve performance and security.

In summary, this study shows how Rust is able to overcome systematic ecosystem problems in bioinformatics, including version mismatches, software distribution and clinical implementation. We also demonstrate how Rust has the potential to become the only language a research group or bioinformatician ever has to master to cover the “full stack” of bioinformatics.

# Materials and Methods

## Choice and use of AI agents

A combination of Claude and Codex was used for this study. As new versions were released during translation, this study should not be seen as a comparison of specific agentic models. In the beginning of the project, it was mainly driven by Claude code (Opus 4.7). Once Codex (20% GPT5.4 and 80% GPT5.5) was deemed more suited for translation work, it became the primary choice, especially during the early critical stages of translation. At the time of writing, GPT-5.5 (Codex) is the default, followed by Claude for resolving difficult tasks. Claude was used for all GUI (graphical user interface) programming, where the screenshots of the original software was provided as references. The translations were produced over a period of 11 weeks, using a Claude Max 20x subscription (11 weeks), and two Codex Pro subscriptions (8 weeks). Up to ~90% of the limit of all three licenses were used during this period.

## Choice of translation targets

The software corpuses included in this translation belong primarily to three categories: (1) Microbial and NGS software that is upstream of the current version of Bascet[19], (2) Software that is of relevance for image analysis and spatial omics, (3) upstream library dependencies, which in particular cover compression and file format support.

## Translation

Naive prompts using Claude to translate software resulted in code that lacked most features, and the replicated features had bugs or gross simplification of algorithms. To get better translation results, a systematic check list of prompts was developed along with translation. The currently recommended prompts are included in Supplemental Material. However, both Codex and Claude have picked up implicit assumptions based on our workspaces, and the prompts may thus not be as efficient for other users. Further independent testing is needed.

The core principles that were used for all later translations (the majority) is that:

- Each original function should be translated into one Rust function

- No additional functions should be introduced
- The logic of each function is independently conserved into the translated function
- The code should follow the original code origanization, including directory structure
- Stubs of all functions and structs are filled in, with idiomatic mapping from original code names to Rust snake_case (this is the Rust naming convention)

Further prompts are injected (see Supplemental material) but these principles ensure (for Codex) that the translation is principled and conservative, enabling later audit.

Most translations were roughly done according to the following idealized steps:

1. Static analysis and scaffolding of translation (i.e., write stubs for all functions)
2. First complete translation of the code
3. Testing of the code on synthetic data (comparing to original code output)
4. Testing of the code on real data (comparing to original code output)
5. Benchmarking and speed optimization
6. Source code specific cleanup
7. Addition of idiomatic Rust API; deciding if some features should be optional (e.g., making it possible to omit command line interface)

Codex was used for the initial translation of all later corpuses as it was found to be more conservative, both in terms of verbatim translation of individual functions, but also in following the instructions. Because initial mistranslation compounds later code design, it is crucial that the first steps are done conservatively. Claude was later adopted to primarily resolve bugs that Codex could not find, or to find optimization opportunities.

## Translation of C code

C-code translated according to our instructions results in non-idiomatic Rust code, where raw libc functions are frequently used. Such code is not safe and Rust cannot guarantee that the translation has no memory referencing errors, free-after-use errors, nor memory leaks. Such code should be made idiomatic after translation, but with care to not reduce speed or increase RSS in the process. The following is an example prompt used to translate jpegxr (31 kLOC):

*/goal Make this code more rust idiomatic. This means data structures, functions etc need to rather use rust typical storage choices. No C types should remain and no C strings should remain. All malloc and free should be replaced by rust idiomatic memory allocation. Replace C strings with vec<u8> if not sure if unicode is safe. replace pointers with rust references when equivalent and performance is not lost(!). Use Option to represent possible null values. Whenever possible, replace data structures that are casted with named enums to avoid any need of casting. Ensure that the exceptions or errors that were originally handled are handled by translation. Do not introduce helper functions but keep translation 1-1 as far as possible. Refactor aggressively. Code need not compile at each step. Do not introduce helpers or shims, replace pointers with references optimistically and clean up later (let the code break).*

Additional prompts were added during and after goal execution, e.g.:

- *I think TRUE and FALSE can be removed mechanically across all files quite easily*
- *I checked the code and it looks like most *mut in function declarations can be replaced with &mut. A lot of errors will appear but they can be handlded in parallel*
- *Relax faithful translation: functions related to memory allocation and deallocation need not be translated 1-1. There should be little need to manually free memory in Rust*

Initial translation to non-idiomatic Rust (tested and benchmarked) took 10 hours and 15M tokens (Codex). Conversion to idiomatic Rust was performed in two sessions; first ~8h with Claude, then 40 min Codex (700k tokens). To verify the API, prompts were given to another agent focused on Bioformats-rs: (paraphased) *“Is the API rust idiomatic”* + *“Refactor the API for your needs. Ensure that the API is general and not only fulfilling the needs of Bioformats”*. Letting agents edit across code bases is however generally not recommended as it can unintentionally collapse API separation.

## Development of static analysis support tools

Early translation attempts were made solely using agentic AI. The following errors in translation occurred commonly:

- Expressions were swapped for no reason, i.e., a<b ⇒ b>a, or branches reordered
- The wrong variable types were used, causing overflows
- Order of operations was changed, causing floating point math to give different answers
- Algorithms were replaced with naive alternatives
- When multiple similar functions were present, the wrong function was called or “reused” for a different purpose
- Helper functions were introduced to try to make an existing function able to also solve other problems, cascading into alternative designs that later could not be extended
- SIMD optimizations were omitted

To mitigate these problems, additional prompts were added to make the agents translate the code more verbatim. However, these prompts were ineffective for Claude, and mostly insufficient for Codex. A concern is that the agentic AI’s are not precise enough in tracking the overall code structure. Tracking code structure is however something that classic algorithms can do efficiently and with high precision.

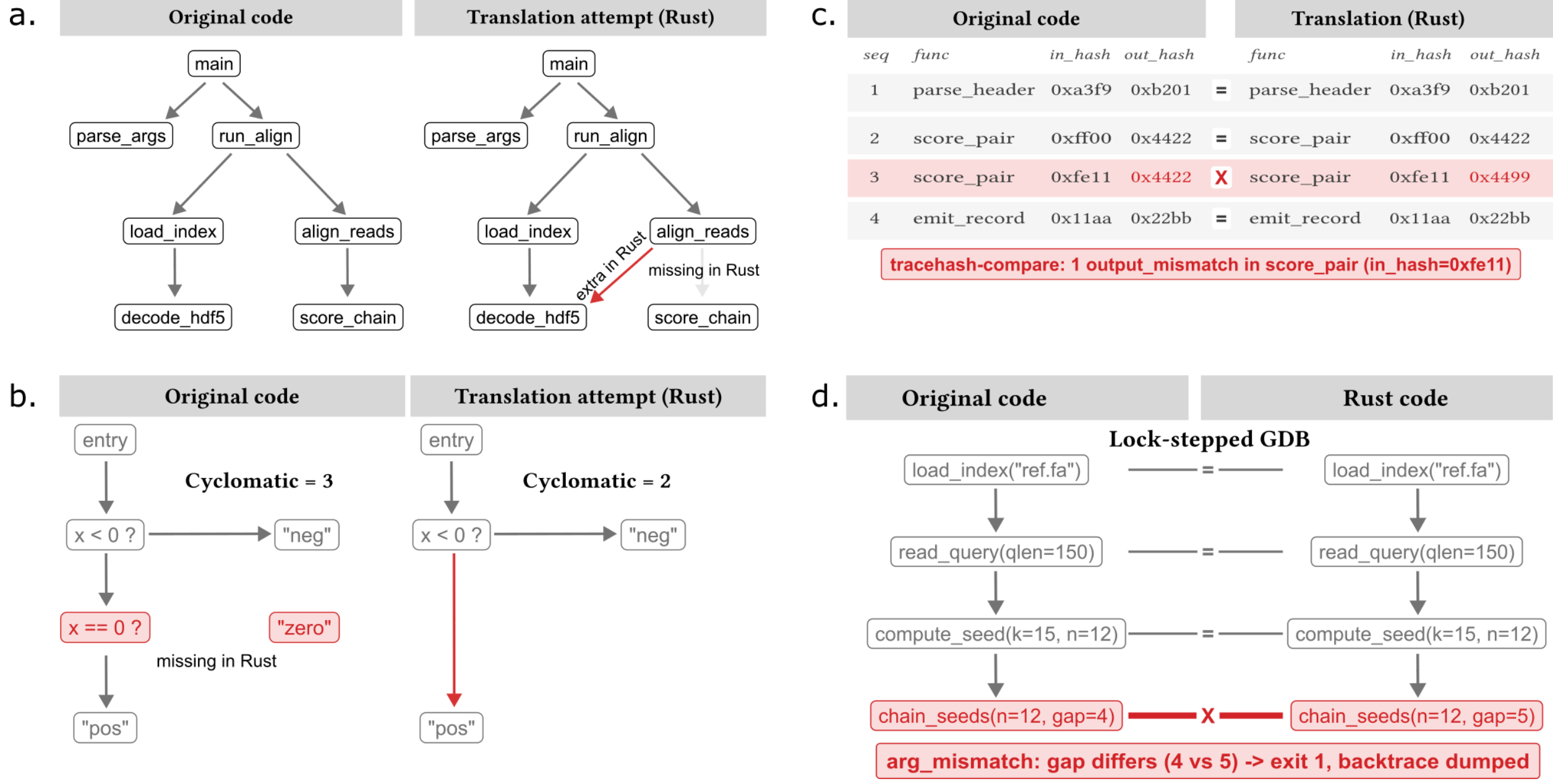

***Figure 1. Development of static analysis support tools. (a)*** *CCC call graph comparison.* ***(b)*** *CCC function complexity comparison.* ***(c)*** *tracehash comparison of functions via instrumentation.* ***(d)*** *gdbtv comparison of functions via locksteppped debuggers.*

To implement such algorithms, static analysis tools were thus developed using Claude ("vibe coded"). The library Treesitter (https://github.com/tree-sitter/) was used for parsing the original code into an AST (abstract syntax tree). This in turn enabled algorithmic analysis of code complexity and call graph structure. The agents are then prompted about their location which contains agent handover instructions, enabling them to figure out how to use the tools. Manual reminders were also prompted whenever the translation progress seemed stuck.

The following tools were developed (**Figure 1**):

- ***CCC*** (Code complexity comparator) compares the call graph of the original code and candidate translation. In addition, it compares functions by the presence of types, symbols (numbers and strings), cyclomatic index[20], and logical expressions (first order logic). CCC relies on the presence of a mapping file (Rust function vs original function) which can be generated by agentic AI, or a first order guess is made. CCC also has a graphical user interface to compare the translated code *vs* the original but it was not extensively used for later translations.

- ***tracehash*** aids code instrumentation. The inputs and outputs of functions are hashed, enabling rapid comparison of deviations (assuming pure functions) and difference in how many times the functions are called (higher level logic failure). *tracehash* also enables output of non-hashed data for selected functions, which can be used to better understand deviations once they have been detected.

- ***gdbtv*** *(gdb-translation-verifier-rs)* compares the original and translated code in parallel GDB sessions. This is similar to the idea of bisimulation, used to prove equivalence of software[21]. Unlike the first two tools, neither Claude nor Codex reached for this tool automatically, suggesting that their training prefers their first two modes of debugging.

To ensure that the tools are useful in practice, the agentic AI (primarily Claude) was instructed to add features as needed (in the original code repository) during initial use of the tools. In particular, the file describing which original function corresponds to which Rust function was improved to handle ambiguous cases; and lifetime functions (constructors and destructors) are given different treatment to relax the 1-1 correspondence (Rust does not typically need destructors, and constructors are implemented differently).

## Validation and use of static analysis support tools

During development, the agents are instructed to use the static analysis tools, and both AI agents invoke them automatically (primarily *CCC* and *tracehash*). However, use of these tools turned out insufficient to remove all translation errors.

Later translation attempts focused on improving translation efficiency. After initial scaffolding, parallel agents can be used to do the translation. The translation is performed bottom-up following the call graph extracted by CCC. Qualitatively, it appears that parallelization also made the AI agents fix translation errors faster. A behavior frequently noted is that, once an error was discovered, the AI agents would give most attention to a smaller set of functions rather than perform a broad audit. This was especially the case when searching for performance problems (i.e. the agent analysis a “hot spot”). However, when the translation failed to progress, the error was typically elsewhere, causing a cascade of problems. Use of parallel subagents enforced a broader search that enabled such problems to be found. As a bonus, by adressing general translation faithfulness instead of one particular error, multiple problems could be fixed in parallel.

While AI agents can easily be prompted to audit code, they will struggle to find all issues if not done systematically on large code bases. Instead, a better strategy is to perform per-file auditing. Agents may also miss problems during one audit session and this must be accounted for. Codex tended to be better at fixing problems and generated less false positives, but both Claude and Codex could be productively used for auditing. The following prompts (paraphrased) were used for systematic audit:

1. *I want to audit each file and fix problems. To ensure all files are checked properly, list all files in TOAUDIT.md along with check boxes. Each file must pass two consecutive audits without complaints before it is considered ok. I want function names to systematically map to Rust snake case. Each original function should have one rust function. Do not add helper functions beyond this. Ensure that all logic is covered and that all functions are retained.*
2. */goal fix all files according to TOAUDIT.md and update file as needed*

For VTK, one of the largest codebases, initially translated by Claude, a systematic parallel audit by Codex took about 4 days. However, systematic audits did not remove all errors, as revealed by testing on real data.

## Benchmarking

All benchmarks were chosen by Claude or Codex, with some degree of user input to steer the choice. The primary prompts for adding benchmarks were analogous to, e.g., *"Compare speed, RSS, parity to original code on realistic data", "Test on larger real data. Do not simply replicate test data we already have"* and *"Test on 100M reads".* Additional prompts include *"ensure that the comparison is fair"* and *"ensure that the JVM has been warmed up before measuring".* Codex was used for the final benchmarking of each program as Claude frequently generated biased benchmarks.

Because each program has a large number of subprograms and parameters, individivudal benchmarks should not be interpreted as definite; rather, their primary purpose has been used for regression (e.g., to check if algorithms have not been translated properly). However, their overall trend across corpuses is interpreted in this study to give a qualitative idea about the potential for speed improvement across different types of source material.

All benchmarks were performed on an Intel Xeon Gold 6138 2.00GHz, 192GB RAM. Different number of thread counts and input data sizes were stressed. Real data was used for most of the final benchmarking.

## Use of agentic AI for manuscript generation

Codex was used to estimate numbers cited in this text, using the following prompts. The numbers must be interpreted qualitatively (typos in prompts included as-is):

- "which software are used for analysis of microbial analysis, covering isolates and metagenomics?" + "give me a tool count, excluding GUIs" → 61
- "I want as estimate of how many packages these tools and libraries need as upstream dependencies. run in parallel" + "I want transitive package count" → "~2,000-3,500 unique transitive software packages, excluding databases."
- "see benchmarks.csv [this is a file listing the github repositories seen in benchmark, Figure 1]. I want to know how many lines of code, excluding unit tests and comments, of rust. developing some utility for this, then use parallel agents" → "700,098 Rust LOC, excluding comments and tests".

Draft figures were generated as vector graphics using Claude, and polished using Inkscape. Agentic AI has not been used to edit the text in any way.

# Results

## Hybrid agentic AI translation is possible for most but not all software

Out of approximately 40 translation attempts, 35 software packages across imaging, NGS (next-generating sequencing) and upstream libraries have been driven to a state where they have been tested, benchmarked, and are ready for use by early adopters (i.e., translation bugs are still expected to be present but we already use the code in production).

The time required to translate a code base varies greatly. The latest translation, jpegxr (C++, 31k lines of code) was translated and verified in less than 20 hours using the latest prompts provided (Materials and Methods; 10% faster without targeted optimization); other programs such as BWAMEM2 required weeks where validation and optimization took most of the time. An important note is that agentic AI by default is single-threaded and does not make efficient use of modern computers. However, the current prompts explicitly instruct the AI agents to parallelize, and this is especially important when using Codex as it is less prone to automatically spawn parallel subagents. Claude spawns parallel agents if the workload is suited. To further parallelize further, the simplest way is to translate multiple codebases in parallel.

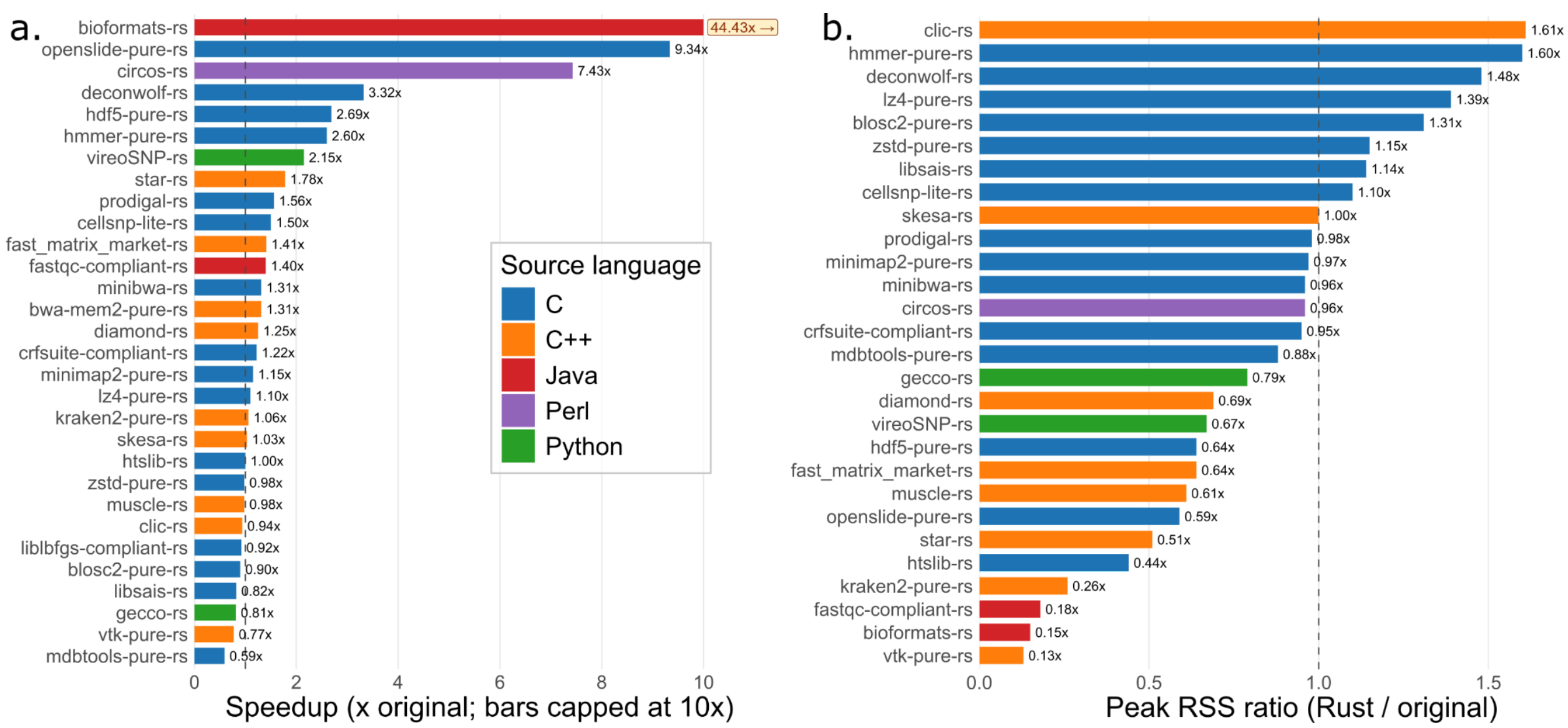

***Fig 2. Best-effort estimates of the Rust-translated versions improvements over the original code*** *in regards of* ***(a)*** *speed and* ***(b)*** *RAM memory usage. The measurements must be interpreted qualitatively as different benchmarks have been applied to each software (see Materials and Methods). However, trends are visible in terms of original language and domain of application.*

All translated packages were optimized but under the constraint that the logic for each function should be retained in its Rust counterpart (1-1 translation). Thus, most optimizations related to improving memory usage, and in particular, to avoid the need to copy data. Common Rust performance tricks are to rather take a reference to a data (instead of a copy), return a reference to a data, or to add data to a preallocated buffer instead of returning a new buffer (useful whenever the buffers can be cleared and reused). Only in rare cases did the translation use the Rust *unsafe* keyword (which enables safety checks to be disabled) and raw pointers (the Rust default is to instead use a safer start-to-end range as a reference, also called a “fat pointer”). The AI agents were unlikely to use these unsafe constructs without active prompting but also had a tendency to needlessly copy memory during the first translation pass. However, both Codex and Claude have added code to “forget” memory, effectively causing memory leaks; thus the user still needs to perform some type of manual audit of the code.

The performance of different translations are compared in **Figure 2**. In most cases, Rust was able to match or exceed the speed of the original software. The time spent optimizing the code was unevenly distributed, with highly optimized software (C/C++) requiring by far more work to reach speed parity (such as the *de novo* assembler SKESA, and various compression libraries). The speedup of other translations vary greatly, and since Python and R libraries both tend to embed C/C++ code, the expected speedup is highly case dependent. Java benefitted from a great speedup but Java is notoriously hard to benchmark due to its design. The Java virtual machine (JVM) interprets code by default (like R/Python) but its hotspot compiler detects code that is run repeatedly and compiles the code on the fly. The code must thus first be “warmed up” to give representative benchmarks. Some care was taken during benchmarking but the bottomline is that Java is not strictly comparable.

The ability to lower or retain memory usage (RSS; resident set size) varies greatly. The fastest software also tends to have the lowest RSS as memory bandwidth is increasingly a bottleneck in modern computers[22]. Dynamically typed languages such as R/Python typically have the highest RSS overhead as they need to track the types of all variables, while the Rust compiler can omit this information during runtime. In several translations, Rust however has higher RSS than the original software. It is likely that all of these cases can be fixed but either requires (1) giving up on the 1-1 translation, or (2) using raw pointers, or (3) using more complex multithreading buffer strategies. Because this affects the possible faithfulness or safety of the code, aggressive attempts at reducing RSS further have not been performed. Finally, Java is again an outlier, where memory is also reserved for the JVM. This memory is somewhat offset for larger workloads but since the JVM both relies on garbage collection and does typing by erasure, memory overhead is largely unavoidable.

Translation of C code is feasible but the output is non-idiomatic Rust. Because the generated code contains a large number of pointers and manual memory management via libc, this code does not capture the safety guarantees of Rust. Such code can however be rewritten to better match Rust idioms using a dedicated translation pass (see Materials and Methods).

Metaprogramming remains a particular challenge for the provided translation approach. Two attempts were made at translating BLAST[23] but neither resulted in a useful product. There are several obstacles to the approach taken in this study related to metaprogramming: (1) BLAST generates source code during the build process, (2) BLAST combines C and C++, (3) BLAST relies heavily on the C++ preprocessor to generate code, (4) BLAST also makes heavy use of templates. The concept of one Rust function per original function breaks down when a large amount of the code revolves around generating functions (i.e., metaprogramming). C++ templates also do not always map to Rust generics, as Rust generic types are required to satisfy certain traits (e.g., additivity), while C++ template types are better thought of as "cut and paste". Rust generics also lack type specialization. More research is thus needed in translation strategies for metaprogramming-heavy code.

## Translation to Rust removes the need for Conda and containers, and increases portability

Our work on translation was prompted by our issues in developing Zorn/Bascet[19] - the first single-cell preprocessing software aimed at microbial metagenomics. Unlike single-cell RNA-seq analysis of eukaryotes, which has a small number of fixed stages[24] (debarcode, align, count features), single-cell microbial analysis is more akin to genome isolate analysis, for which a large number of packages exist. We aimed to integrate them for single-cell needs but largely failed to deliver a product using established methodology, i.e., packages via Conda, wrapped in a container (Docker or apptainer). The failure was due to a mix of (1) Zorn/Bascet being a complex package and workflow manager, (2) Conda failing to resolve compatible packages, (3) Conda failing to resolve versions in reasonable time, (4) many configurations of containers being required to cover all operating systems and CPU architectures. All of these problems could be resolved by translation.

By translating all upstream dependencies, version management is now solely managed by the Rust package manager Cargo. Rust and Cargo are designed to enable conflicting versions of software to be installed in parallel, unlike R, Python and most other programming languages. This removes the need and issues related to the Conda package manager.

After translation, all upstream software of Bascet are now used as libraries, removing the fragile CLI interface in favour of statically typed Rust functions. But by using the code as libraries, only a single binary is produced, containing the whole ecosystem (Bascet + upstream software; **Figure 3a**). This removes the need for containers to glue the files together. Thus, we now distribute naked binaries (200mb - 80x smaller than the original container). Together with the removal of Conda, compilation time is also reduced from 20min to about 1-2 min.

Cross-compilation also enabled distribution of native binaries for Apple M3 computers, increasing performance on non-Intel hardware (containers will require hardware emulation). Finally, during translation, we also conditionally disabled or replaced Unix/Posix-specific instructions (e.g., setting of file permissions, posix threads, etc). This has made Bascet the first single-cell toolkit that can run under native Windows (i.e., no need to install WSL2 subsystem).

## Translation improves performance primarily for linked libraries

Single-cell analysis requires the processing of a large number of inputs. Modern single cell datasets can encompass up to 100M cells[25]. This amount of data requires highly efficient software and storing the data of each cell in an individual file is not feasible due to file system design[19]. Furthermore, running command line utility on each file would have major overhead. Assuming one second/cell in startup time, this would amount to a total of 1,157 days of just software starting cost. However, many programs load databases (from HMM profiles to entire KMER databases), resulting in loading times up to minutes. It is thus not reasonable to merely call other programs using standard procedures (such as in Nextflow[9] or Snakemake[26] pipelines) in this context.

To remove the software startup overhead, we added high-level APIs to each package, and divided it into separate steps (**Figure 3b**): (1) input file parsing and loading, (2) database loading, (3) processing, and (4) saving output. Out of these, only processing must be done for each cell. Furthermore, copying of memory (from one variable to another) is an increasingly expensive operation. By adapting the API to operate on input data by reference, zero-copy strategies can sometimes be achieved. Taken together, software such as SKESA[27] and GECCO[28] run over 3 times faster after integration in Bascet[19] (benchmarks part of manuscript in preparation).

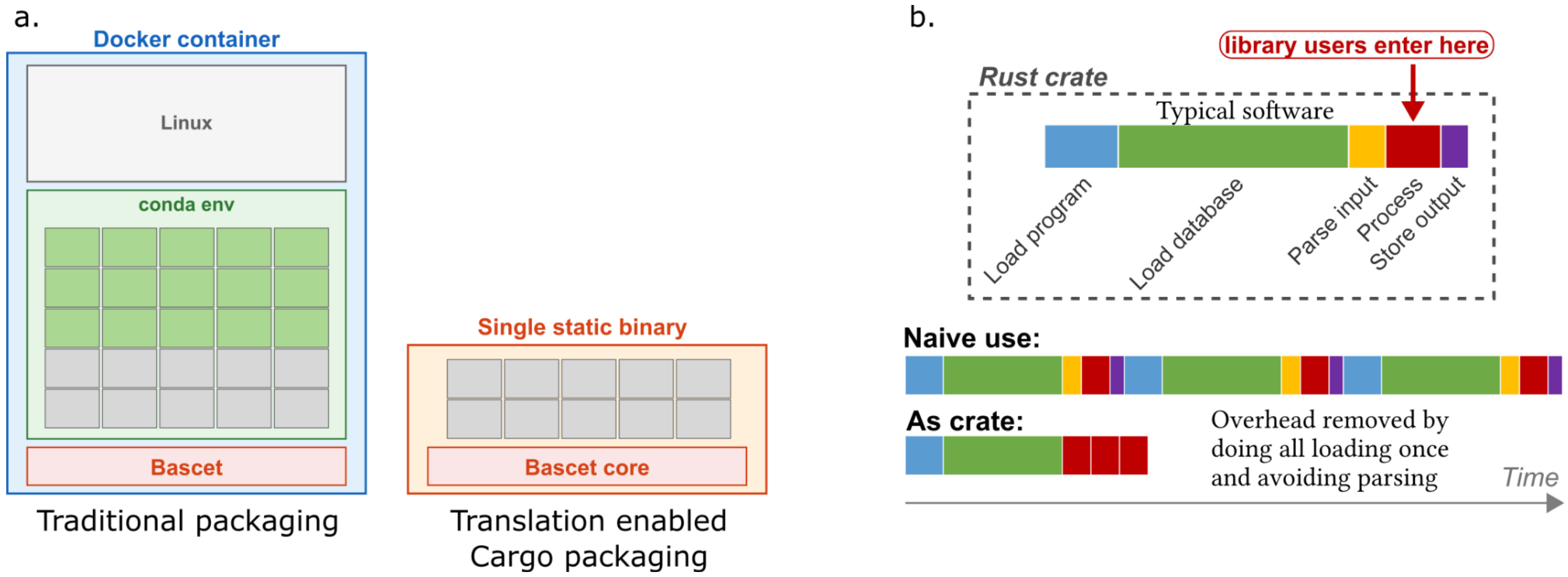

***Fig 3. Packaging of translated dependencies via Cargo. (a)*** *Translation enables all dependencies to be packaged in a single Rust executable, removing the need for Conda and containers.* ***(b)*** *Direct access to translated functions unlocks further optimization by omitting unnecessary data transfer and library initialization.*

## Floating point handling is hazardous

Unlike in theoretical math, floating operations are not commutative, i.e., (ab)c ≠ a(bc). This made translation of some software (such as HMMER[29]) especially difficult as the AI agents frequently did not respect order of operations. Handling of integer types (and over/underflow) was also a problem, but generally easier to work out.

The reliance on precise floating point combined with imprecise translation resulted in p-values differing to some degree from the original code. This in turn caused problems during validation, as the small errors escalated to larger ones: (1) Order of results, such as gene lists, being different. (2) Results being omitted due to p-value cut-offs.

All floating point errors are resolved in the translation but certain types of optimizations could not be performed; e.g., SIMD (Single Instruction Multiple Data) are specialized CPU instructions that can greatly improve speed for certain workloads (e.g., vector multiplications). They however reorder the operations, resulting in slightly different results for floating point math. A future possibility is to add SIMD as an optional feature, but with a warning that the results will differ from the original software.

# Discussion

This study shows that large-scale translation of open source code is possible using a combination of suitable prompts, static analysis support software, and agentic AIs. This will have particular impact on research software, which has higher technical debt than other open source software. This can be expected, given its higher degree of complexity, combined with a smaller user base. Appropriate use of agentic AI can thus help catch up with the technical debt.

This study focuses on Rust as a possible single language for all bioinformatics applications. We show that performance is comparable to C/C++, and that code written in Java, Perl and Python can be translated with great speed improvements. The memory usage can also be reduced, but idiomatic Rust can also penalize memory usage, likely due to its stricter ownership model. Reducing memory further may require larger rewrites of the code, or the use of unsafe language features.

We show that use of Rust as the single language removes the need for Conda and container infrastructure, removing a major layer of complexity, and largely resolves the issue of how to handle conflicting software versions. The size of distributed software can be greatly reduced. Rust cross-compilation also supports other CPU architectures such as the Apple M3.

A major concern is the extent to which translation introduces bugs in the code. This does happen, and a new cycle of software testing will be required. However, a large number of bugs can be found by simply comparing the output of the translated software *vs* original code output, on a large number of varied corpuses of real data. This study found it essential to use real world data, and that the tests generated by agentic AIs are largely insufficient. A challenge is that complex software has many settings or algorithms, making comprehensive testing difficult. Because of the 1-1 correspondence during translation (one function in, one function out), it should however be possible to design specialized testing software. Some of the functions are, e.g., similar enough that even mathematical equivalence proofing could be a route forward[21]. It also suggests that specialized translation algorithms (i.e. transpilers) might be able to handle a large amount of translation, while also reducing cost of AI and speeding up translation.

During this translation, at least 3 bugs were uncovered, suggesting that the bugs introduced may be offset by the bugs being resolved. The bugs involved use of a broken dependency (bug since long already resolved but new version not integrated), use-after-free (possible memory corruption) and an unhandled corner case. Further memory related bugs may have been resolved implicitly by removing pointer-based memory management (malloc/free). A fair assumption is that the agentic AI technology will keep maturing to the point where bugs as a whole will be net negative.

In the case of Bascet, we directly interfaced upstream software through Rust functions. Besides possibly massive speed increases, this is the only way that safety can be guaranteed *across* software. Current bioinformatics practice frequently makes use of untyped Unix pipes, but this approach does not guarantee that the data is compatible across software. The translation approach removes the need for pipes, and named pipes, neither of which are compatible with Windows. Native Windows compatibility is increasingly important as many users now use Ubuntu-in-windows (WSL2), making it impossible to run Linux commands from RStudio.

"Every line of code is a liability". By maintaining translations rather than producing greenfield implementations, the increase in total software maintenance burden is limited. In addition, provenance is clarified, addressing the major concern about the copyright of AI-generated code. During this study, the translations for two programs (openslide, minibwa) were made to track newer versions. The 1-1 translation ensured that upstream changes were localized to well-defined spots in the translation, enabling updates taking less than one hour. However, many upstream changes were related to the build system, documentation, and other aspects not affecting the translation. This suggests that a large number of repositories could be maintained with little effort using parallel AI agents. However, this would put the bioinformatics ecosystem at risk of supply-chain attacks, as has plagued the Javascript community (via the Node package manager[30]). A more conservative approach to updating may thus be suitable.

The translations also highlighted the challenges in precise p-value reproduction. This will become an even larger problem as bioinformatics code starts to use GPUs, as for the sake of performance, some operations are not guaranteed to occur in deterministic order. A partial solution is to ban to avoid or ban p-value cutoffs, in favour of probabilistic weighing schemes that are less sensitive to rounding errors. As p-value cutoff tuning enables p-hacking[31], and handling of larger data has become easier, the field is overdue in taking this step.

Finally, while this study shows the feasibility of translation, there is still ample room for better translators and verifiers. Translation guidelines have already been generated (e.g. https://rewrites.bio/), but the sudden appearance of the AI technology calls for a rethinking of bioinformatics. The ability to refactor software at massive scale enables strategic opportunities for better integration of our many software packages, resolving big challenges: Compute needs can be reduced, barriers to adoption removed, environmental impact reduced and the reliability of software increased - which is especially important as NGS is adapted in the clinic. As such, this work hopefully inspires greater future endeavours in bioinformatics.

# Code availability

Each translated code base, and static analysis tools, is available in its own repository (**Table 1**). Benchmark data has been deposited at https://github.com/henriksson-lab/paper_rust_translation1. The latest updates on our translation work is available at https://github.com/henriksson-lab/rustification.

# Competing interests

The authors declare no conflict of interest

# Acknowledgements

This work was supported by: the Swedish Research Council (VR; grants 2024-03952, 2024-06085); the Swedish Foundation for Strategic Research (SSF; grant ITM24-0035); Cancerfonden (grant #23 3102 Pj); Computation was enabled by resources provided by High Performance Computing Center North (HPC2N; Umeå University, Umeå, Sweden).

Thanks are given to all users who provided bug reports and test data during early translation attempts. Particular thanks are given to Heng Li for testing, input and benchmarking.

# Supplemental Material

| Category | Link and reference |
|---|---|
| **Static-analysis and audit tools** | Code-complexity-comparator / CCC; Tracehash-rs; gdb-translation-verifier-rs / gdbtv |
| **Alignment, search, and MSA** | blast-rs[32]; diamond-rs[33]; bwa-mem2-pure-rs[34]; minimap2-pure-rs[35]; star-rs[36]; muscle-rs [37]; hmmer-pure-rs [29] |
| **Annotation, variants, assembly, and QC** | kraken2-pure-rs[38]; prodigal-rs[39]; prokka-rs[40]; gecco-rs[28]; skesa-rs[27]; vireoSNP-rs[41]; cellsnp-lite-rs[42]; fastqc-compliant-rs (bioinformatics.babraham.ac.uk/projects/fastqc/); ncbi-amrfinderplus-rs[43] |
| **Visualization and file/image I/O** | circos-rs[44]; bioformats-rs[10]; htslib-rs[45]; vtk-pure-rs (vtk.org); openslide-pure-rs[46] |
| **Microscopy, EM, and ML components** | micromanager-rs[47]; clic-rs [48]; itk-rs[49]; cellpose-rs[50]; deconwolf-rs[51]; YOLOv11-rs[52] |
| **Compression, formats, numerics, and ML primitives** | hdf5-pure-rs (hdfgroup.org/solutions/hdf5); blosc2-pure-rs (blosc.org); lz4-pure-rs (lz4.org); zstd-pure-rs[53]; libsais-rs[54]; fast_matrix_market-rs (github.com/alugowski/fast_matrix_market); crfsuite-compliant-rs (chokkan.org/software/crfsuite); liblbfgs-compliant-rs (github.com/chokkan/liblbfgs) |

***Table 1:*** *Developed static analysis software, and translated software*

# Guide to AI-mediated translation of code to Rust

By Johan Henriksson, https://www.henlab.org/

# The overall translation cycle

1. Git clone the original code in a sub directory. Then git init . in current directory
2. **Check if the code makes sense to translate at this point**. Good judgement needed!
    a. This guide assumes that the original code is well structured and does not do too much crazy pointer hacking. Otherwise CCC might struggle to deliver
3. Set up a translation scaffold i.e. stubs to fill in. The better the foundation is, the better is the result! Consider nuking if the foundation is poor and you are struggling
4. Ask to get a README.md. Edit manually. Ensure that people know that this is an LLM translation, and double check if the license is set correctly. Add citation information. Add git hash to track what was translated
5. Fill in all the stubs bottom up. The LLM adds a lot of tests automatically but they are of limited value
6. Add the CLI layer (typically use the “clap” crate)
7. Perform testing on real data
8. Fix problems. Fill in gaps that were missed
9. Fix speed and RSS. Rust translations tend to .clone() a lot. This increases memory and kills performance! *Treat differences in RSS and speed as a regression - if your code is too fast, maybe it doesn’t do all that it should?*
10. Remove code that is not portable
11. Add a high-level API

Below are prompts in blue to get you going. **This is a starting point only. You will need to adapt the prompts for your needs.**

# Prompts for Claude

I have not been able to make Claude follow the prompts in this document, or a longer list of explicit prompts. You can use Claude to resolve problems that Codex cannot find, because all LLMs can end up stuck on problems and needing a new perspective. Changing LLM sometimes works. But by all means, do NOT use Claude for early stages of translation, as early mistranslations propagate throughout the entire design. If this happens, restarting the translation will likely be faster than trying to refactor it to the correct shape.

# Codex prompts: Translation of Python/R code

Time can be saved if an analysis is performed on the input file data types initially. This can be fixed later otherwise.

- If you have code that generates the input file, point Codex to analyze that code
- Otherwise make qualified guess from looking at the input files

# Codex prompts: Before translation

What are the dependencies of this project? What Rust crates can we replace them with?
Back out of big translations until you have dependencies covered! Get experience with small translations first.

# Codex prompts: New translation

The aim is to create a faithful (identical output) rust translation of code into a rust crate. original code in a subdir. Ideally each original function becomes one rust function to help audit (don't add helper functions). New source files should reflect original code organization (but e.g. xxx.c and xxx.h should be merged). Code should also follow the same directory structure. Names should be mapped systematically to Rust snake case. When translating more code, aim to translate bottom up and include all logic in the first pass, to avoid later backtracking and debugging. If original code uses templates, we should likely aim to use equivalent generics. Prioritize real world data when testing the code.

Treat helper functions as bugs, as they should not be needed.

Before starting, Study the following crates for use during translation:
Remember: code-complexity-comparator is at /data/henriksson/github/claude/code-complexity-comparator .

Start by creating a scaffold including all structs. Generate one rust stub function for each input function (remove dead code warnings). Set up ccc_mapping.toml as soon as possible. CCC can be used to see which order to implement functions (i.e. bottom up).

Then create README.md. The readme should contain the git hash of the code we are translating. Copy any LICENSE file. Add citation information to the readme.

Make a plan and save it in TODO.md

**If you think it is needed you can also add the text below. You can also add it later if debugging is hard:**

Remember: tracehash is at https://crates.io/crates/tracehash-rs . Also study /home/mahogny/github/claude/gdb-translation-verifier-rs

**Notes**:
- More conservative translation in the beginning wins big time. It avoids tracing errors at the end, which is the most time consuming part otherwise
- Both Claude and Codex have a tendency to generate "adapter functions". Some type of helper functions to feed data. But these make audit really difficult so it is best to ensure you don't get them in the first place. CCC does not know how to handle #define macros, so just inline everything for the sake of clarity (some codes better than others in this regard)

# Codex prompts: Continuing a part-translated project

The aim is to create a faithful (identical output) rust translation of code into a rust crate. original code in a subdir. Ideally each original function becomes one rust function to help audit (don't add helper functions). New source files should reflect original code organization (but e.g. xxx.c and xxx.h should be merged). Code should also follow the same directory structure. Names should be mapped systematically to Rust snake case. When translating more code, aim to translate bottom up and include all logic in the first pass, to avoid later backtracking and debugging. If original code uses templates, we should likely aim to use equivalent generics. Prioritize real world data when testing the code.

Treat helper functions as bugs, as they should not be needed.

Before starting, Study the following crates for use during translation:
Remember: code-complexity-comparator is at /data/henriksson/github/claude/code-complexity-comparator .

Make a plan and save it in TODO.md

**Some work was already done. Figure out what and then continue**

**If you think it is needed you can also add the text below. You can also add it later if debugging is hard:**
Remember: tracehash is at https://crates.io/crates/tracehash-rs . also study /home/mahogny/github/claude/gdb-translation-verifier-rs

**Notes:**
- Almost the same as before. Copy-paste of all but with the addition to try and pick the pace. Not sure if needed. Try to never end a conversation to keep life simple (use screen or tmux). Compacting conversations may help in different ways (less token use, better focus), but I never do this (let me know your experience)

# Codex prompts: Continuous initial work

/goal Finish items in TODO.md and update the file as you go. Use parallel workers when it makes sense. Note that function calls follow a DAG. There can be multiple files at the bottom, enabling parallelism
To reduce the amount of prompting, I am now keeping a list of things to do in TODO.md. The /goal command then lets you track it to the end.

What is left to do?
Get a suggestion on what to work on. Hopefully something useful, but this can end up being a slur of trivial tests if used too often

Add above to TODO.md
One of many ways to add suggested items to TODO.md. But be prepared to steer it manually if needed

# Codex prompts: Midway / parity fixing

Once the LLM mainly seems to be adding trivial tests, you need to push it harder

What real datasets can we test this software on?
It is key to use real data. If it takes real data and repeats it 10 times then that is a bad test

Fix / fix all
Inevitably, you will find issues. You get far just prompting “fix”

**If the code is slow, test on really small data, then move to perform speed optimization for a while before coming back here. It is possible that algorithms have not been translated well, or that the translated code copies too much memory. It is easier to expose this problem during optimization**

Audit the code for missing logic
This reminds the LLM to check the original source code instead of trying to reason about why there are errors. Needed at times, especially if it is stuck debugging function X when the error is in Y.

Audit translation of types, especially f32 vs f64, broadly. Do in parallel.
Use CCC binary operations metric.
Floating point math is especially tricky to get right. CCC binary operations metric counts use of * etc, and can be used to flag functions where math has been simplified.

Ensure there is one rust function per original input function. Do not add helper functions. Ensure there is one xxx.rs for each xxx.c and xxx.h, and that original directory structure is mirrored. Do in parallel
The 1-1 translation tends to glide; be sure to bring it back once in a while

Fix all in parallel. One worker per file. Each function should mirror its own original function
This prompt and variations can be needed to reinforce the 1-1 translation. Codex might otherwise try to compensate for the error in one function by adding errors to another, creating a mess

it is fine to break the code for a while. refactor aggresively
*Essential prompt if the agent doesn't progress*. It prioritizes compiling code at each step but this can make it get stuck during refactoring. Commit code first

# Codex prompts: Performance

**Always prioritize correctness over speed. Don't bother with speed until code is at least somewhat correct**

Compare the speed and RSS with the original code
The translation always starts out slower than the original code, especially if C/C++.

Improve speed, focus on weak spots, but ensure faithful translation
The LLM might try to make the code faster for the best case unless told otherwise. Also, don't let it start adding heuristics to improve speed

Audit code usage to bring down RSS to original code
The #1 reason Rust might be slower than the original is that it clones data all over. RSS is a good diagnostic for this

Audit the code for missing logic
The #2 reason Rust might be slower than the original is that it failed to capture some logic. It might be working on the wrong thing!

Is there any SIMD we have not yet translated? Is it all wired in?
I haven't used this prompt, but the LLM is not keen on using SIMD, and it typically does not get translated in the first pass. It might need some forcing as Codex otherwise starts doing microoptimizations before considering SIMD translation.

Audit for out-of-bounds checks. Are there any that we can eliminate? Do we need to use pointers or unsafe?
Absolutely last resort.

Perf rust and original code to find differences
This is one way of guiding the LLM to optimize the right place instead of inventing new optimizations.

# General things to do to check performance

Not everything can be described as prompts. You will need to take a different approach depending on your software

- Test using both 1 thread, and 20-40+ threads. Both are extreme cases, exposing (1) rayon overhead and (2) if I/O suddenly becomes a bottleneck.
- Rayon may need to do “chunks” to reduce overhead. Chunk size is a mess to optimize
- Buffers may need to be reused to avoid mallocs
- Sometimes you can reduce the number of mallocs by keeping multiple objects together. This starts to go away from faithful translation though (unless already done)

# Codex prompts: Post-translation/compatibility

Does this code use non-portable API calls? Such as the use of unix file descriptors. If so, suggest portable fixes
Legacy code bases might use posix calls and other things. There is little need for that these days, and it prevents compilation on Windows (not that anybody likes Windows, but someone is likely forced to use it).

# Codex prompts: Post-translation/Docs

Carry over and adapt docstrings for each function from the original code
(could maybe add right away in first stage, or get CCC to do)

# Codex prompts: Post-translation/pointer abuse

Are there pointers used that we can remove without affecting speed?
I have had rare cases of pointers being kept for no good reason (C code). They also get introduced during optimization if one forces it. This prompt is not typically needed

# Codex prompts: Clean-up

Scan the code for #[allow(dead_code)]. Investigate each use. Remove #[allow(dead_code)] when not needed. Investigate truly dead code and suggest what to do These are sometimes kept from the first stub creation. Some functions are not used in the end, but that's still worth checking

Set the name of the crate to xxxx-rs

A good name is needed. See also what is free on [crate.io](crate.io)

If you fork the original software:

1. You should probably pick a different name to avoid confusion
2. Consider having a chat with the original software developer about what their thoughts are. Open source licenses will always allow you to do a fork, but it's probably good courtesy to at least inform them prior
3. Ensure you still cite original authors as the source
4. Add a remark that this is not the original software, and that behavior can be expected to deviate

Set the version of the crate to 0.1.0

I recommend keeping all translations to 0.x.y. Don't up the major number unless you are the author of the original software

Remove current benchmarks on README.md. Compare speed and RSS with original code and add to readme. Include ratio. Ensure fair comparison

Benchmarks easily get out of date. Ensure that benchmarks are on real data of sensible size. This prompt is just a starter. Good judgement needed

Prep cargo publish; ensure include lines in cargo.toml are updated

Should always be the final prompt before publishing a crate. May need to run multiple times, after a git commit / push. The include-lines refer to what files should go into the crate. [Crate.io](Crate.io) has an upload limit. Smaller is better!

**Some test data is too large to upload. Make it optional. There is no single simple prompt to take care of this**

What to git add?

Don't add more than needed. But see if the LLM has created unexpected dependencies (e.g., test data in the original source code directory, which I don't check in. copy it out in that case)

Update gitignore

You don't want uncommitted files, or cargo publish will complain

# High-level API

Does the crate have a high-level API?

Leading question. Probably not if it is a regular program. If you translate C code, you will likely need to do extensive rewriting to polish the API

Add a high-level API
This is a start but… **Check the resulting API yourself. Would you want to use it? The LLM can not replace common sense! C-code especially requires polishing**

Use the builder pattern
This is a common way to avoid massive structs of settings. Good for APIs

Can the database be loaded once and used multiple times?
Example of how to see if the API is general enough - separation of starting phase and run phase

Can reads be fed into this crate using Noodles?
Example of how to see if the API is general enough

# Before "cargo publish"

**If you are not a Rust developer, ask for input before publishing. You can host a crate on github, and people can use it. Don't stress to "finish"**

Check if README.md is correct
But you have to check it yourself too. Delete stuff that is not needed

Add examples of CLI usage to README.md
Always useful

Add examples of how to use this crate as a library to README.md
Always useful

It should be possible to use this crate as a library. Make the CLI an optional feature. But keep the clap parser for library use
This one makes it easier to access the CLI if needed. E.g., another program that embeds the software can also add a direct interface. This can be useful to test exactly the same version of the translated crate, but also to expose other functions (Bascet exposes the "index" function of aligners, as an example)

# Test test test

Add more real data tests. Ensure to stress all CLI options
*There is no such thing as too much testing. The LLM is very fallible. You need to attack it from many angles, and in quantity.*

# Systematic audit

I find systematic audit to be needed for larger code bases (i.e. most code of relevance). It is essential if you also use Claude to do any translation work. The idea is to ensure that all files get enough attention. *Italics: Add a targeted reminder as in italics. You can expand beyond this.*

I want to audit each file and fix problems. To ensure all files are checked properly, list all files in TOAUDIT.md along with check boxes. Each file must pass two consecutive audits without complaints before it is considered ok. *I want function names to systematically map to Rust snake case. Each original function should have one rust function. Do not add helper functions beyond this. Ensure that all logic is covered and that all functions are retained.*

Then
Audit/fix files following TOAUDIT.md using parallel workers
or to keep the agent going for a longer time:
/goal fix all files according to TOAUDIT.md and update file as needed

Note that if the problems are deep, you may have to skip "parallel workers" for some passes. But I think Codex figures this out on its own. If it finds gaps you may have to explicitly prompt to fix them

# Combined use with Claude

Claude appears to still be useful during the debug phase, especially to conserve tokens. The "/loop" feature can be nice for running debugging overnight. There are also some low-risk tasks for making documentation.

Definitely add this before any work with Claude:
Never git commit or publish crates without my permission

Examples (not systematically evaluated yet):

- /loop Continue finding and fixing bugs. translate original logic faithfully. improve speed when logic is proven equivalent and original code is faster. Prioritize correctness over speed. Never do hacks, always do the right fix even if it requires more work. Don't stop this loop until bugs fixed and speed on parity
- Carry over and adapt docstrings for each function from the original code. Use parallel subagents
- Carry over and adapt docstrings for each function from the original code. Make a docstring if the original function doesn't have any. Use parallel subagents
- Add citation information to README.md

# Prompts being tested (Codex)

audit code broadly. What is left to do for this translation? add to TODO.md. do in parallel
Combined with
use parallel workers to take care of items in TODO.md. remove items as you go

Possible single prompt
keep testing and fixing, adding missing functionality from TODO.md, using parallel workers. add new items to TODO.md as needed
or
use parallel workers to take care of items on TODO.md. remove items as you go. if TODO.md empty, do parallel audit to add to list

The two-part prompt might be better at ensuring Codex does not get stuck doing trivial fixes

# Translating and cleaning C (not C++) code

Translated C code will likely contain many C'isms that need specific prompts for cleaning up. **Not all C codes can be translated well. Some code bases do not map well to Rust 1-1 translations. You can do the initial steps but be prepared to retreat if the code is unsuited for 1-1 translation.**

1. First, you can use the recipe above to make a first pass translation. The main challenge is that the code will be filled with the use of malloc/callocs/free (manual memory allocation), C-style strings (null-terminated), libc calls, etc. For C, it is also unknown how long arrays are - this information has to be passed around separately (and there is no distinction between a list and a point to a regular variable).
2. Then the real challenge begins. It will not be possible to refactor individual functions because they are all interlocked in their representation. Claude can absolutely help with this refactoring because it can be such a mess. Here is a prompt:

Make a plan to refactor this code to be idiomatic Rust. This means that no calls to libc are performed, repr["C"] is not used, FFI is not used, memory allocation is not handled via malloc/calloc/free. Strings are not stored as Cstr but instead use Vec<u8> (we will evaluate if unicode can be used later; be conservative with strings for now). Pointers should not be used, but rather turn all pointers into references. The code need not compile for many cycles until the refactoring is complete. Don't add helper functions or shims during the refactoring as these will regress the auditabilty of the code. The ABI need not be C-compatible (only Rust code will call this crate). Replace C-types such as c_int, c_char etc with Rust types.

Key points here:

- Explicitly request that the code need not compile as neither Codex/Claude will automatically take this step. Forcing compilability in each step really locks down what changes can be made to the code
- Neither LLM is happy to drop repr[“C”] without explicit instruction. The source of pointers is the data structures and getting rid of them is priority

You will spend plenty of time refactoring toward the goal above. At some point you will need to take a bigger step where you need to use your judgement:

Do not use void*. Instead, we should either use dyn Box traits, or named enums. Audit the code for use of void* and suggest the most appropriate Rust type instead. Named enums are a good default

In my experience, it is not common that you need dyn box traits. Named enums allow “match” expressions and are better. Void* is primarily used to store different kinds of data, with the type decided by another variable. This is exactly what a named enum is.

**This is not a full list of cleaning that you need to do.** You will definitely need to read the generated Rust code and think how you would want to rewrite it